\documentclass[conference]{IEEEtran}

\usepackage{algorithmic}
\usepackage{amsmath,amssymb,amsfonts}
\usepackage{balance}
\usepackage{booktabs}
\usepackage[strict]{changepage}
\usepackage{cite}
\usepackage{colortbl}
\usepackage[T1]{fontenc}
\usepackage{framed}
\usepackage[acronym, shortcuts]{glossaries-extra}
\usepackage{graphicx}
\usepackage[hidelinks]{hyperref}
\usepackage{multirow}
\usepackage{newtxmath}
\usepackage{siunitx}
\usepackage{textcomp}
\usepackage{xcolor}
\usepackage{url}
\usepackage{svg}

\usepackage{subcaption}

\def\BibTeX{{\rm B\kern-.05em{\sc i\kern-.025em b}\kern-.08em
    T\kern-.1667em\lower.7ex\hbox{E}\kern-.125emX}}

\definecolor{formalshade}{rgb}{0.85,1,0.85}
\definecolor{darkblue}{rgb}{0.0,0.6,0.30}
\newenvironment{formal}{%
  \MakeFramed{\advance\hsize-\width\FrameRestore}%
  \noindent\hspace{-4.55pt}%
  \begin{adjustwidth}{}{7pt}%
}
{%
  \end{adjustwidth}\endMakeFramed%
}

\glssetcategoryattribute{acronym}{nohyper}{true}
\setabbreviationstyle[acronym]{long-short}

\newacronym{can}{CAN}{Controller Area Network}
\newacronym{dsa}{DSA}{Digital Signature Algorithm}
\newacronym{ecu}{ECU}{Electronic Control Unit}
\newacronym{iot}{IoT}{Internet of Things}
\newacronym{kem}{KEM}{Key Encapsulation Mechanism}
\newacronym{nist}{NIST}{National Institue of Standards and Technology}
\newacronym{ntru}{NTRU}{N\textsuperscript{th} degree Truncated polynomial Ring Units}
\newacronym{mlwe}{MLWE}{Module Learning-With-Errors}
\newacronym{pqc}{PQC}{Post-Quantum Cryptography}

\def\Sphincs+{SPHINCS\textsuperscript{+}}

\makeatletter
\newcommand{\linebreakand}{%
  \end{@IEEEauthorhalign}
  \hfill\mbox{}\par
  \mbox{}\hfill\begin{@IEEEauthorhalign}
}
\makeatother
    
\begin{document}

\title{PQ-CAN: A Framework for Simulating Post-Quantum Cryptography in Embedded Systems}

\author{\IEEEauthorblockN{Mauro Conti}
\IEEEauthorblockA{\textit{Department of Mathematics} \\
\textit{University of Padova}\\
Padova, Italy \\
mauro.conti@unipd.it}
\and
\IEEEauthorblockN{Francesco Marchiori}
\IEEEauthorblockA{\textit{Department of Mathematics} \\
\textit{University of Padova}\\
Padova, Italy \\
francesco.marchiori@math.unipd.it}
\and
\IEEEauthorblockN{Sebastiano Matarazzo}
\IEEEauthorblockA{\textit{Department of Mathematics} \\
\textit{University of Padova}\\
Padova, Italy \\
sebastiano.matarazzo@studenti.unipd.it}
\linebreakand 
\IEEEauthorblockN{Marco Rubin}
\IEEEauthorblockA{\textit{Department of Mathematics} \\
\textit{University of Padova}\\
Padova, Italy \\
marco.rubin.5@studenti.unipd.it}}

\maketitle

\begin{abstract}
The rapid development of quantum computers threatens traditional cryptographic schemes, prompting the need for Post-Quantum Cryptography (PQC).
Although the NIST standardization process has accelerated the development of such algorithms, their application in resource-constrained environments such as embedded systems remains a challenge.
Automotive systems relying on the Controller Area Network (CAN) bus for communication are particularly vulnerable due to their limited computational capabilities, high traffic, and need for real-time response.
These constraints raise concerns about the feasibility of implementing PQC in automotive environments, where legacy hardware and bit rate limitations must also be considered.

In this paper, we introduce PQ-CAN, a modular framework for simulating the performance and overhead of PQC algorithms in embedded systems.
We consider the automotive domain as our case study, testing a variety of PQC schemes under different scenarios.
Our simulation enables the adjustment of embedded system computational capabilities and CAN bus bit rate constraints.
We also provide insights into the trade-offs involved by analyzing each algorithm's security level and overhead for key encapsulation and digital signature.
By evaluating the performance of these algorithms, we provide insights into their feasibility and identify the strengths and limitations of PQC in securing automotive communications in the post-quantum era.
\end{abstract}


\section{Introduction}
\label{sec:introduction}

The rapid progress of quantum computing technologies presents a profound threat to traditional cryptographic systems at the backbone of modern communications.
As quantum computers advance, they threaten the security of widely used cryptographic algorithms, particularly those based on integer factorization~\cite{rivest1978method} and discrete logarithms~\cite{diffie1976}, which are vulnerable to Shor's algorithm~\cite{shor1999polynomial}.
Symmetric cryptographic systems, on the other hand, are affected by Grover's algorithm~\cite{grover1996fast}
In response, \ac{pqc} has emerged as a crucial field, offering algorithms designed to resist quantum-based attacks by relying on computationally hard mathematical problems, such as lattice-based, code-based, and multivariate polynomial challenges.
The importance of \ac{pqc} has been further emphasized by the ongoing \ac{nist} standardization process, which aims to establish cryptographic algorithms that can withstand quantum attacks while remaining practical for real-world deployment~\cite{nist2024fips}.

Despite the progress in \ac{pqc} standardization, integrating these algorithms into resource-constrained environments, such as embedded systems, remains a significant challenge.
Indeed, embedded devices, including those used in industrial control systems, \ac{iot}, and automotive applications, often have strict limitations on computational power, memory, and energy consumption~\cite{fernandez2019pre}.
Instead, \ac{pqc} algorithms demand substantial computational resources, increased memory for key storage, and higher bandwidth for transmitting larger cryptographic primitives, posing integration challenges in such constrained environments.
One critical domain where these constraints are particularly evident is the automotive industry, where the \ac{can} bus serves as the backbone of in-vehicle communication.
Designed for efficiency and reliability, the CAN bus lacks built-in cryptographic protections, making it vulnerable to conventional and quantum-era attacks~\cite{canbus_standard}.
Implementing \ac{pqc} in this environment requires careful consideration of algorithmic overhead, latency, and real-time constraints to ensure secure and efficient communication between embedded systems in automotive electronics, called \acp{ecu}.
Understanding the trade-offs between security and performance is essential for assessing the feasibility of \ac{pqc} in securing automotive networks against future threats.
This raises a critical question: \textit{can \ac{pqc} be implemented in embedded systems with varying constraints?}
In particular, we identify the following research questions.

\begin{enumerate}
    \renewcommand{\labelenumi}{\textbf{RQ\theenumi}}
    \item \label{rq1} Can \ac{pqc} be implemented in existing lower-end embedded devices?
    \item \label{rq2} Can \ac{pqc} be implemented in time-sensitive applications?
    \item \label{rq3} What is the trade-off between computational overhead and security level with \ac{pqc} in embedded systems?
\end{enumerate}

\textit{Contribution.}
In this paper, we present \textbf{PQ-CAN}, a modular and comprehensive framework designed to evaluate the feasibility of \ac{pqc} algorithms in embedded systems.
Using the automotive domain as a case study, we analyze several \ac{pqc} algorithms under varying computational capabilities to assess their practical application.
Our findings highlight a critical trade-off between security strength and computational overhead, demonstrating the challenges of implementing specific algorithms in legacy embedded systems.
Built with scalability in mind, PQ-CAN is implemented in Docker, allowing it to simulate a wide range of embedded systems not limited to the automotive domain.
Our contributions can be summarized as follows.
\begin{itemize}
    \item We introduce PQ-CAN, a novel, modular framework for evaluating \ac{pqc} algorithms in embedded systems.
    \item We present a detailed study of \ac{pqc} algorithms in the automotive domain, offering insights into their feasibility under different system constraints.
    In particular, we focus on schemes pertaining to \acp{kem} and \acp{dsa}.
    \item We make PQ-CAN open-source, providing public access at: \texttt{\url{https://github.com/spritz-group/PQ-CAN}}.
\end{itemize}

\textit{Organization.}
The remainder of this paper is organized as follows.
Section~\ref{sec:related} reviews background knowledge and related works on \ac{pqc} and automotive systems.
We report our methodology in Section~\ref{sec:methodology}, and we show the results of our evaluation in Section~\ref{sec:evaluation}.
Finally, we provide our discussion in Section~\ref{sec:discussion}, and Section~\ref{sec:conclusions} concludes our work.
\section{Related Works}
\label{sec:related}

In this section, we review related works on \ac{pqc} (Section~\ref{subsec:pqc}) and \ac{can} bus security (Section~\ref{subsec:can}).

\subsection{Post-Quantum Cryptography}
\label{subsec:pqc}

\ac{pqc} aims to secure communications against quantum attacks, primarily using lattice-based, code-based, and multivariate polynomial cryptography.
Lattice-based schemes such as CRYSTALS-\textsc{Kyber}~\cite{bos2018crystals} and CRYSTALS-Dilithium~\cite{ducas2018crystals} are favored for their efficiency and security, while \textsc{Falcon}~\cite{prest2020Falcon} and \Sphincs+~\cite{bernstein2019sphincs} offer alternative signature mechanisms with varying trade-offs.
However, these algorithms demand significant computational resources, making them challenging for resource-constrained environments like embedded systems.
To address this, research has focused on optimizing these schemes for lightweight applications~\cite{liu2024post}.
Still, balancing security and computational overhead remains a challenge~\cite{kuznetsov2023trade}.

\subsection{CAN Bus}
\label{subsec:can}

The \ac{can} bus is a multi-master, message-based communication protocol developed by Robert Bosch GmbH in the early 1980s~\cite{canbus_standard}.
It was designed to exchange data efficiently between \acp{ecu}, and due to its fault tolerance, high reliability, and lightweight communication overhead, \ac{can} has become a widely adopted standard across multiple industries, including automotive.
A \ac{can} network uses a two-wire differential signaling scheme for electromagnetic interference resistance.
It operates at the OSI data-link layer, employing a frame-based structure with fields like an identifier, data payload, CRC, and acknowledgment bits, supporting data rates up to \SI{1}{Mbps}.
\ac{can} enables broadcast communication between \acp{ecu}, which manage most vehicle functions, ranging from critical real-time operations to infotainment, and can vary significantly in computational capabilities based on their role.
Despite its efficiency, \ac{can} lacks fundamental security mechanisms such as authentication, encryption, and access control~\cite{avatefipour2018state}.
As a result, research has increasingly focused on implementing authentication mechanisms to enhance its security~\cite{10735339}.
\section{Methodology}
\label{sec:methodology}

In this section, we report our proposed methodology for simulating \ac{pqc} algorithms in embedded systems.
We overview our simulation environment in Section~\ref{subsec:simulation}, and we discuss the considered algorithms in Section~\ref{subsec:cryptographic}.

\subsection{Simulation}
\label{subsec:simulation}

Our simulation framework provides a flexible and scalable environment for evaluating \ac{pqc} in embedded systems.
It is built on a containerized architecture, representing each embedded device as an isolated Docker container.
This approach allows for the simulation of different network topologies, hardware constraints, and communication protocols, making it adaptable to various embedded systems beyond the automotive sector.
To accurately model embedded constraints, our framework supports:
\begin{itemize}
    \item CPU frequency constraints to emulate different processing capabilities.
    \item Bit rate limitations to impose real-world communication delays.
    \item Traffic injection mechanisms to simulate network congestion and interference.
\end{itemize}

For our study, we apply this framework to an automotive \ac{can} bus 2.0 environment to analyze the feasibility of \ac{pqc} in vehicle networks.
Each \ac{ecu} container runs an isolated cryptographic process in this context, with all containers sharing the same network namespace.
Among these, Alice and Bob represent the two \ac{ecu} implementing the cryptographic communication.
The implementations adhere closely to \ac{nist}'s C language API conventions,\footnote{\url{https://csrc.nist.gov/projects/post-quantum-cryptography/pqc-archive}} with only minor deviations.
We then develop different programs for Alice and Bob implementing cryptographic schemes and linking them to a separate cryptographic library optimized with AVX-2 instructions.
The communication happens on a virtual \ac{can} bus, whose interface is offered by the \texttt{vxcan} driver, which creates two ends of a communication link (one in the host, the other in the network namespace of the containers).
The host can set the bit rate for both ends, e.g., via the \texttt{tc} utility of the \texttt{iproute2} suite, and the Linux kernel will manage the traffic so as not to exceed the requested limit.
Hyper-Threading and Turbo Boost are disabled, and two cores are reserved for Alice and Bob through the \texttt{taskset} utility from the \texttt{util-linux} package.
We also fix the clock frequency of these cores via Docker's \texttt{cpus} flag, which constrains the containers to a maximum usage of the host's CPU cycles (i.e., by setting it to the fraction of the target frequency over the host frequency, as $\texttt{target\_freq} / \texttt{host\_freq}$).

\subsection{Cryptographic Algorithms}
\label{subsec:cryptographic}

\begin{figure}[!htbp]
    \begin{subfigure}{0.495\columnwidth}
        \includegraphics[width=\columnwidth]{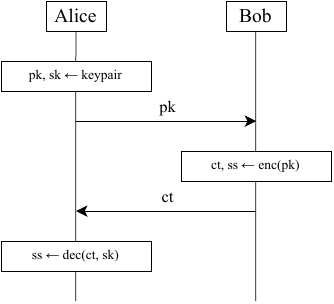}
        \caption{KEM scheme.}
        \label{fig:kem}
    \end{subfigure}
    \hfill
    \begin{subfigure}{0.495\columnwidth}
        \includegraphics[width=\columnwidth]{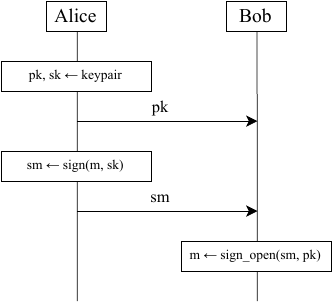}
        \caption{DSA scheme.}
        \label{fig:dsa}
    \end{subfigure}
    \caption{Diagrams of the cryptosystems considered in this study.}
    \label{fig:schemes}
\end{figure}

This work evaluates seven post-quantum cryptographic schemes, focusing on \ac{kem} (Fig.~\ref{fig:kem}) and \ac{dsa} (Fig.~\ref{fig:dsa}) algorithms.
For \ac{kem}, we consider CRYSTALS-\textsc{Kyber}, BIKE, HQC, and Classic McEliece in Section~\ref{subsub:kem}.
For \ac{dsa}, we consider CRYSTALS-Dilithium, \Sphincs+, and \textsc{Falcon} in Section~\ref{subsub:dsa}.

\subsubsection{KEM Algorithms}
\label{subsub:kem}

These algorithms enable secure key exchange over insecure channels by allowing two parties to establish a shared secret without directly transmitting the key.
The process involves encapsulating the secret using a public key and decapsulating it with a private key, ensuring confidentiality for encrypted communication.

\paragraph{CRYSTALS-\textsc{Kyber}}
A lattice-based \ac{kem} that relies on the \ac{mlwe} problem~\cite{bos2018crystals}.
It balances security and efficiency with smaller key sizes and faster computations than other lattice-based schemes.
\textsc{Kyber} comes in three variants: \textsc{Kyber}512, 768, and 1024, each providing different trade-offs between security and performance.

\paragraph{BIKE}
A code-based \ac{kem} relying on the Quasi-Cyclic Moderate-Density Parity-Check (QC-MDPC) code problem, offering security against quantum attacks through hard-to-decode random linear codes~\cite{BIKE2024}.
BIKE comes in three variants: BIKE Level-1, Level-3, and Level-5, which balance security and performance, with BIKE Level-1 being the most efficient and BIKE Level-5 offering the highest security at a higher computational cost.

\paragraph{HQC}
A code-based \ac{kem} that secures key exchange using the Syndrome Decoding Problem (SDP) in the Hamming metric~\cite{HQC2024}.
It leverages Quasi-Cyclic (QC) linear codes to balance security and efficiency, though its key and ciphertext sizes are larger than lattice-based schemes.
HQC offers three variants: hqc-128, hqc-192, and hqc-256---providing increasing levels of security at the cost of higher computational and storage requirements.

\paragraph{Classic McEliece}
A code-based \ac{kem} relying on the hardness of decoding random binary Goppa codes~\cite{bernstein2017classic}.
It offers strong security, but suffers from extremely large public key sizes and challenging deployment in constrained environments.
Classic McEliece provides multiple parameter sets, with higher values offering increased security at the cost of larger keys and computational overhead.
``f'' variants exist for faster key generation without altering security parameters.

\subsubsection{DSA Algorithms}
\label{subsub:dsa}

These algorithms ensure message authenticity and integrity through verifiable signatures, allowing a sender to sign messages with a private key and a receiver to verify them using the corresponding public key.

\paragraph{CRYSTALS-Dilithium}
A lattice-based \ac{dsa} securing signatures through the MLWE and Module Short Integer Solution (MSIS) problems~\cite{ducas2018crystals}.
It uses rejection sampling to prevent side-channel leaks and operates with integer-based arithmetic for efficiency.
Dilithium offers three variants: Dilithium Level 2, Level 3, and Level 5---balancing security and performance, with higher levels providing stronger protection at the cost of larger keys and signatures.

\paragraph{\Sphincs+}
A hash-based \ac{dsa} offering post-quantum security without algebraic assumptions, relying solely on cryptographic hash functions~\cite{bernstein2019sphincs}.
It uses Merkle trees, one-time signatures, and few-time signatures, ensuring robustness but with larger signatures and higher computational costs.
\Sphincs+ supports SHA-256, SHAKE256, and Haraka, with security levels of 128, 192, and 256 bits.
It offers ``fast'' and ``small'' optimizations, as well as ``robust'' and ``simple'' variants, balancing security and efficiency.

\paragraph{\textsc{Falcon}}
A lattice-based \ac{dsa} based on the Short Integer Solution (SIS) problem over \ac{ntru} lattices, offering compact signatures with strong security~\cite{prest2020Falcon}.
It uses Gaussian sampling via FFT for efficient signing but requires careful implementation due to floating-point dependencies.
\textsc{Falcon}-512 (NIST level 1) provides small signatures and fast verification, while \textsc{Falcon}-1024 (NIST level 5) enhances security with larger keys and signatures, balancing efficiency and robustness.
\section{Evaluation}
\label{sec:evaluation}

We now report the results of our experiments.
We first overview our experimental setup in Section~\ref{subsec:setup}, and we disclose our simulation results for \acp{kem} and \acp{dsa} in Section~\ref{subsec:kemresults} and Section~\ref{subsec:dsaresults}, respectively.

\subsection{Experimental Setup}
\label{subsec:setup}

The experimental setup employed in this study consists of a simulated CAN bus 2.0 environment running on Arch Linux 6.12.10, powered by an Intel Core i7-1065G7, with \SI{16}{\giga\byte} of RAM.
To represent a range of different automotive systems, we simulated three \ac{ecu} configurations: a low-end, a mid-range, and a high-end one, selecting their clock frequencies based on an analysis of the hardware available on the market \cite{ecu1,ecu2}.
The configurations operate with \acp{ecu} running at \SI{120}{\mega\hertz} (\textit{``low''}), \SI{200}{\mega\hertz} (\textit{``mid''}), and \SI{300}{\mega\hertz} (\textit{``high''}), each connected to a \ac{can} bus supporting a maximum bit rate of \SI{125}{Kbps}, \SI{500}{Kbps}, and \SI{1}{Mbps}, respectively.
Since Alice and Bob are not synchronized, message loss is possible, reflecting real-world \ac{can} bus scenarios where \acp{ecu} operate asynchronously.
This allows us to assess the success rate of a plain implementation of the schemes in Fig.~\ref{fig:schemes}.
Therefore, a 2-second timeout was set on the receiver side, mimicking practical timeout mechanisms used in automotive networks to prevent indefinite waiting.
To simulate background traffic, we implement another container, Charlie, loading the CAN bus to 88\% capacity with the \texttt{canbusgen} utility of the \texttt{can-utils} suite.
The sender has no timeout, replicating cases where \acp{ecu} experience transmission delays under heavy traffic.
For each algorithm and configuration, we perform 100 iterations, measuring execution times, success rates, and overhead.
Outliers due to processor frequency instability were omitted.
Under heavy traffic, the long receiver timeout sometimes led to delayed packets, increasing success rates by extending the reception window.
Overhead includes cryptographic operations and memory management.
For \acp{kem}, it covers key generation to decapsulation, while for \acp{dsa}, we isolated cryptographic overhead by subtracting nominal message transmission time from total execution time.

\subsection{KEM Results}
\label{subsec:kemresults}

Our analysis, represented in Table~\ref{tab:kem}, shows that communication overhead dominates total execution time, with cryptographic operations accounting for only about 13\% of it on average.
Among the tested algorithms, \textsc{Kyber}512 consistently exhibits the lowest overhead and fastest cryptographic operations across all configurations.
As security levels increase, the other \textsc{Kyber} variants maintain a strong balance between performance and security, outperforming alternatives with lower security levels.
Among level 1 security algorithms, BIKE Level-1 is significantly slower than even the most secure \textsc{Kyber} variant, \textsc{Kyber}1024, while hqc-128 already lags behind other contenders.
Higher security levels exacerbate these delays: BIKE Level-3, BIKE Level-5, hqc-192, and hqc-256 exhibit substantial execution times, with the latter reaching up to 5 seconds in the ``low'' performance configuration.
Interestingly, security level alone does not dictate performance; public key and ciphertext sizes significantly impact communication times.
More insight on this is shown in our repository.
However, within each algorithm family, higher security variants generally follow expected trends in execution time.
Success rates tend to drop in lower-performance configurations.
Still, algorithms with longer communication times, such as hqc-192 and hqc-256, show higher success rates due to the extended reception window, as explained earlier.
In contrast, faster algorithms complete transmission too quickly to benefit from this effect.
We excluded Classic McEliece from our comparison due to its excessive latency, which ranged from 8 to 49 seconds even in its ``f'' variants under the high configuration.
The variants of \textsc{Kyber} remain the best choice for optimal balance between speed and security, delivering execution times at least 300 times shorter than hqc-192 and hqc-256 while maintaining an acceptable success rate.

\begin{table*}[!htbp]
    \renewcommand{\arraystretch}{1.25}
    \caption{Comparison of the \ac{kem} schemes sorted by the lowest overhead.}
    \label{tab:kem}
    \centering
    \begin{tabular}{l|c|c|c|c|c|c|c}
        \hline
        \multirow{2}{*}{\textbf{Algorithm}} & \multirow{2}{*}{\textbf{Set}} & \multicolumn{4}{c|}{\textbf{Timings [\si{\milli\second}]}} & \multirow{2}{*}{\bfseries \shortstack{Success\\Rate}} & \multirow{2}{*}{\bfseries \shortstack{Security\\Level}} \\
        \cline{3-6} & & \textbf{Key Generation} & \textbf{Encapsulation} & \textbf{Decapsulation} & \textbf{Overhead $\downarrow$} & & \\ \hline \rowcolor{gray!15}
                                     & high & $0.045 \pm 0.019$  & $0.063 \pm 0.024$  & $0.030  \pm 0.018$  & $1.189    \pm 0.583$  & $\mathbf{0.93}$  &                      \\ \rowcolor{gray!15}
                                     & mid  & $0.048 \pm 0.014$  & $0.064 \pm 0.011$  & $0.032  \pm 0.009$  & $1.080    \pm 0.193$  & $0.82$           &                      \\ \rowcolor{gray!15}
        \multirow{-3}{*}{\textsc{Kyber}512}   & low  & $0.042 \pm 0.019$  & $0.055 \pm 0.023$  & $0.028  \pm 0.013$  & $1.126    \pm 0.628$  & $0.59$           &  \multirow{-3}{*}{1} \\ \hline
                                     & high & $0.049 \pm 0.018$  & $0.070 \pm 0.029$  & $0.038  \pm 0.020$  & $1.459    \pm 0.593$  & $\mathbf{0.94}$  &                      \\
                                     & mid  & $0.059 \pm 0.018$  & $0.074 \pm 0.015$  & $0.044  \pm 0.013$  & $1.457    \pm 0.305$  & $0.77$           &                      \\
        \multirow{-3}{*}{\textsc{Kyber}768}   & low  & $0.052 \pm 0.021$  & $0.068 \pm 0.028$  & $0.039  \pm 0.016$  & $1.356    \pm 0.343$  & $0.64$           &  \multirow{-3}{*}{3} \\ \hline \rowcolor{gray!15}
                                     & high & $0.073 \pm 0.065$  & $0.086 \pm 0.058$  & $0.061  \pm 0.037$  & $1.998    \pm 0.749$  & $\mathbf{0.90}$  &                      \\        \rowcolor{gray!15}
                                     & mid  & $0.065 \pm 0.021$  & $0.091 \pm 0.017$  & $0.055  \pm 0.015$  & $1.825    \pm 0.377$  & $0.80$           &                      \\        \rowcolor{gray!15}
        \multirow{-3}{*}{\textsc{Kyber}1024}  & low  & $0.060 \pm 0.029$  & $0.078 \pm 0.029$  & $0.050  \pm 0.025$  & $1.682    \pm 0.450$  & $0.67$           &  \multirow{-3}{*}{5} \\ \hline
                                     & high & $0.363 \pm 0.188$  & $0.141 \pm 0.054$  & $1.291  \pm 0.934$  & $3.542    \pm 1.466$  & $\mathbf{0.91}$  &                      \\
                                     & mid  & $0.455 \pm 0.129$  & $0.157 \pm 0.032$  & $1.333  \pm 0.329$  & $3.583    \pm 0.624$  & $0.89$           &                      \\
        \multirow{-3}{*}{BIKE Level-1}    & low  & $0.345 \pm 0.116$  & $0.126 \pm 0.049$  & $1.237  \pm 0.706$  & $3.242    \pm 1.113$  & $0.59$           &  \multirow{-3}{*}{1} \\ \hline \rowcolor{gray!15}
                                     & high & $0.073 \pm 0.038$  & $0.201 \pm 0.045$  & $0.354  \pm 0.084$  & $137.374  \pm 4.645$  & $\mathbf{0.96}$  &                      \\        \rowcolor{gray!15}
                                     & mid  & $0.097 \pm 0.027$  & $0.211 \pm 0.010$  & $0.316  \pm 0.024$  & $272.631  \pm 9.025$  & $0.74$           &                      \\        \rowcolor{gray!15}
        \multirow{-3}{*}{hqc-128}    & low  & $0.071 \pm 0.033$  & $0.177 \pm 0.060$  & $0.291  \pm 0.112$  & $1099.459 \pm 40.068$ & $0.72$           &  \multirow{-3}{*}{1} \\ \hline
                                     & high & $0.975 \pm 0.480$  & $0.309 \pm 0.075$  & $4.523  \pm 1.187$  & $140.559  \pm 5.243$  & $\mathbf{0.92}$  &                      \\
                                     & mid  & $1.223 \pm 0.363$  & $0.322 \pm 0.011$  & $4.516  \pm 0.651$  & $274.841  \pm 9.259$  & $0.74$           &                      \\
        \multirow{-3}{*}{BIKE Level-3}    & low  & $0.867 \pm 0.403$  & $0.260 \pm 0.085$  & $4.420  \pm 1.618$  & $1080.292 \pm 39.313$ & $0.65$           &  \multirow{-3}{*}{3} \\ \hline \rowcolor{gray!15}
                                     & high & $2.291 \pm 1.113$  & $0.605 \pm 0.141$  & $9.702  \pm 3.206$  & $258.926  \pm 11.190$ & $\mathbf{0.96}$  &                      \\        \rowcolor{gray!15}
                                     & mid  & $2.977 \pm 0.883$  & $0.608 \pm 0.012$  & $11.408 \pm 1.508$  & $501.347  \pm 9.840$  & $0.88$           &                      \\        \rowcolor{gray!15}
        \multirow{-3}{*}{BIKE Level-5}    & low  & $2.184 \pm 0.975$  & $0.500 \pm 0.171$  & $10.393 \pm 9.710$  & $1961.925 \pm 38.717$ & $0.58$           &  \multirow{-3}{*}{5} \\ \hline
                                     & high & $0.159 \pm 0.095$  & $0.413 \pm 0.113$  & $0.699  \pm 0.274$  & $358.374  \pm 14.621$ & $\mathbf{0.97}$  &                      \\
                                     & mid  & $0.175 \pm 0.059$  & $0.422 \pm 0.015$  & $0.626  \pm 0.036$  & $717.210  \pm 15.302$ & $0.80$           &                      \\
        \multirow{-3}{*}{hqc-192}    & low  & $0.140 \pm 0.061$  & $0.341 \pm 0.129$  & $0.551  \pm 0.287$  & $2728.391 \pm 37.556$ & $0.74$           &  \multirow{-3}{*}{3} \\ \hline \rowcolor{gray!15}
                                     & high & $0.291 \pm 0.142$  & $0.700 \pm 0.231$  & $1.312  \pm 0.456$  & $598.037  \pm 8.012$  & $\mathbf{0.99}$  &                      \\ \rowcolor{gray!15}
                                     & mid  & $0.352 \pm 0.078$  & $0.785 \pm 0.026$  & $1.192  \pm 0.031$  & $1234.307 \pm 14.672$ & $\mathbf{0.90}$  &                      \\ \rowcolor{gray!15}
        \multirow{-3}{*}{hqc-256}    & low  & $0.248 \pm 0.109$  & $0.598 \pm 0.225$  & $1.045  \pm 0.435$  & $4921.836 \pm 32.078$ & $0.72$           &  \multirow{-3}{*}{5} \\ \hline
    \end{tabular}
\end{table*}

\subsection{DSA Results}
\label{subsec:dsaresults}

In our analysis, Alice generated a random 32-byte message using NIST's \texttt{randombytes} function at each run, with its execution time included in the overhead.
In Table \ref{tab:dsa}, only the NIST-standardized \Sphincs+ variants are considered, omitting the “robust” and “Haraka” versions.
The “s” variants and those offering security level 5 were also excluded due to poor performance: the former had near-zero success rates, as Alice’s signing time exceeded Bob’s 2-second timeout. At the same time, the latter achieved at most a 2\% success rate.
The analysis shows that in \textsc{Falcon}-512 and \textsc{Falcon}-1024, where public key and signature sizes are relatively small, key generation dominates the overhead.
Communication time becomes the primary contributor as key and signature sizes grow, particularly in \Sphincs+, where long signatures significantly increase signing times.
Despite their higher processing times, Dilithium and \Sphincs+ suffer from lower success rates than \textsc{Falcon} due to synchronization issues.
Indeed, Bob often fails to resume listening in time, missing Alice’s signed message and triggering the timeout.
Among the schemes, \textsc{Falcon}-1024 has slightly higher overhead than Dilithium Level 5 in ``high'' and ``mid'' configurations but outperforms all Dilithium variants in ``low''.
It also achieves the highest success rates in ``mid'' and ``low,'' though in ``high'', \textsc{Falcon}-512 performs marginally better.
The \Sphincs+ variants exhibit the worst success rates and longest overheads, with some reaching nearly 10 seconds in the ``low'' configuration.
\textsc{Falcon}-512 offers the best balance of low overhead and high success rate, though it remains slower than the nominal baseline.
Dilithium’s lower success rates make it less ideal if stronger security is needed, while \textsc{Falcon}-1024 provides both high security and success rates at the cost of more significant overhead.

\begin{table*}[!htbp]
    \renewcommand{\arraystretch}{1.25}
    \caption{Comparison of the \ac{dsa} schemes sorted by the lowest overhead.}
    \label{tab:dsa}
    \centering
    \begin{tabular}{l|c|c|c|c|c|c|c|c}
        \hline
        \multirow{2}{*}{\textbf{Algorithm}} & \multirow{2}{*}{\textbf{Set}} & \multicolumn{5}{c|}{\textbf{Timings [\si{\milli\second}]}} & \multirow{2}{*}{\bfseries \shortstack{Success\\Rate}} & \multirow{2}{*}{\bfseries \shortstack{Security\\Level}} \\
        \cline{3-7} & & \textbf{Nominal} & \textbf{Key Generation} & \textbf{Signing} & \textbf{Verification} & \textbf{Overhead $\downarrow$} & & \\ \hline \rowcolor{gray!15}
                                                             & high & $0.180 \pm 0.070$ & $43.037  \pm 39.540$  & $0.492   \pm 0.237$   & $0.089 \pm 0.088$ & $46.995   \pm 40.303$  & $\mathbf{0.96}$ &                     \\        \rowcolor{gray!15}
                                                             & mid  & $0.178 \pm 0.070$ & $76.429  \pm 53.586$  & $0.606   \pm 0.222$   & $0.112 \pm 0.080$ & $90.810   \pm 52.814$  & $0.84$          &                     \\        \rowcolor{gray!15}
        \multirow{-3}{*}{\textsc{Falcon}-512}                         & low  & $0.209 \pm 0.115$ & $139.859 \pm 77.103$  & $0.528   \pm 0.243$   & $0.123 \pm 0.092$ & $167.530  \pm 85.312$  & $0.75$          & \multirow{-3}{*}{1} \\ \hline \rowcolor{white}
                                                             & high & $0.180 \pm 0.070$ & $0.071   \pm 0.032$   & $0.144   \pm 0.094$   & $0.117 \pm 0.034$ & $61.275   \pm 7.689$   & $0.28$          &                     \\        \rowcolor{white}
                                                             & mid  & $0.178 \pm 0.070$ & $0.078   \pm 0.026$   & $0.154   \pm 0.085$   & $0.111 \pm 0.011$ & $120.438  \pm 9.534$   & $0.35$          &                     \\        \rowcolor{white}
        \multirow{-3}{*}{Dilithium Level 2}                       & low  & $0.209 \pm 0.115$ & $0.068   \pm 0.025$   & $0.133   \pm 0.092$   & $0.090 \pm 0.026$ & $470.414  \pm 38.111$  & $0.29$          & \multirow{-3}{*}{2} \\ \hline \rowcolor{gray!15}
                                                             & high & $0.180 \pm 0.070$ & $0.099   \pm 0.045$   & $0.338   \pm 0.181$   & $0.158 \pm 0.047$ & $101.937  \pm 5.303$   & $0.50$          &                     \\        \rowcolor{gray!15}
                                                             & mid  & $0.178 \pm 0.070$ & $0.124   \pm 0.039$   & $0.312   \pm 0.150$   & $0.154 \pm 0.010$ & $201.297  \pm 9.269$   & $0.61$          &                     \\        \rowcolor{gray!15}
        \multirow{-3}{*}{Dilithium Level 3}                       & low  & $0.209 \pm 0.115$ & $0.092   \pm 0.042$   & $0.241   \pm 0.117$   & $0.143 \pm 0.075$ & $821.569  \pm 48.606$  & $0.50$          & \multirow{-3}{*}{3} \\ \hline \rowcolor{white}
                                                             & high & $0.180 \pm 0.070$ & $0.135   \pm 0.066$   & $0.311   \pm 0.169$   & $0.206 \pm 0.076$ & $142.395  \pm 5.226$   & $0.54$          &                     \\        \rowcolor{white}
                                                             & mid  & $0.178 \pm 0.070$ & $0.184   \pm 0.053$   & $0.405   \pm 0.129$   & $0.219 \pm 0.016$ & $283.322  \pm 10.458$  & $0.51$          &                     \\        \rowcolor{white}
        \multirow{-3}{*}{Dilithium Level 5}                       & low  & $0.209 \pm 0.115$ & $0.143   \pm 0.066$   & $0.315   \pm 0.166$   & $0.186 \pm 0.062$ & $1131.983 \pm 38.160$  & $0.43$          & \multirow{-3}{*}{5} \\ \hline \rowcolor{gray!15}
                                                             & high & $0.180 \pm 0.070$ & $138.089 \pm 106.749$ & $1.015   \pm 0.491$   & $0.093 \pm 0.140$ & $159.200  \pm 118.779$ & $\mathbf{0.93}$ &                     \\        \rowcolor{gray!15}
                                                             & mid  & $0.178 \pm 0.070$ & $265.043 \pm 89.260$  & $1.445   \pm 0.198$   & $0.171 \pm 0.156$ & $293.600  \pm 123.264$ & $\mathbf{0.96}$ &                     \\        \rowcolor{gray!15}
        \multirow{-3}{*}{\textsc{Falcon}-1024}                        & low  & $0.209 \pm 0.115$ & $361.655 \pm 164.442$ & $0.921   \pm 0.493$   & $0.251 \pm 0.171$ & $388.039  \pm 173.958$ & $\mathbf{0.92}$ & \multirow{-3}{*}{5} \\ \hline \rowcolor{white}
                                                             & high & $0.180 \pm 0.070$ & $0.784   \pm 0.363$   & $85.286  \pm 58.384$  & $1.800 \pm 0.506$ & $580.442  \pm 64.255$  & $0.21$          &                     \\        \rowcolor{white} 
                                                             & mid  & $0.178 \pm 0.070$ & $1.018   \pm 0.271$   & $134.187 \pm 66.447$  & $1.977 \pm 0.240$ & $1119.833 \pm 66.947$  & $0.27$          &                     \\        \rowcolor{white}
        \multirow{-3}{*}{\shortstack{\Sphincs+-128f\\(SHA2)}}  & low  & $0.209 \pm 0.115$ & $0.898   \pm 0.644$   & $185.383 \pm 81.223$  & $1.709 \pm 0.672$ & $4078.748 \pm 103.034$ & $0.26$          & \multirow{-3}{*}{1} \\ \hline \rowcolor{gray!15}
                                                             & high & $0.180 \pm 0.070$ & $0.820   \pm 0.378$   & $99.156  \pm 63.127$  & $2.036 \pm 0.318$ & $591.207  \pm 52.683$  & $0.14$          &                     \\        \rowcolor{gray!15}
                                                             & mid  & $0.178 \pm 0.070$ & $1.041   \pm 0.301$   & $144.316 \pm 65.958$  & $1.803 \pm 0.064$ & $1131.293 \pm 68.882$  & $0.23$          &                     \\        \rowcolor{gray!15}
        \multirow{-3}{*}{\shortstack{\Sphincs+-128f\\(SHAKE)}} & low  & $0.209 \pm 0.115$ & $0.740   \pm 0.377$   & $163.746 \pm 69.818$  & $1.549 \pm 0.794$ & $4023.570 \pm 91.915$  & $0.18$          & \multirow{-3}{*}{1} \\ \hline \rowcolor{white}
                                                             & high & $0.180 \pm 0.070$ & $1.086   \pm 0.501$   & $104.257 \pm 61.055$  & $3.149 \pm 1.157$ & $1196.703 \pm 23.207$  & $0.04$          &                     \\        \rowcolor{white}
                                                             & mid  & $0.178 \pm 0.070$ & $1.412   \pm 0.421$   & $234.477 \pm 72.334$  & $2.964 \pm 0.261$ & $2432.741 \pm 42.550$  & $0.09$          &                     \\        \rowcolor{white}
        \multirow{-3}{*}{\shortstack{\Sphincs+-192f\\(SHA2)}}  & low  & $0.209 \pm 0.115$ & $1.060   \pm 0.503$   & $281.709 \pm 114.678$ & $2.413 \pm 0.719$ & $9066.902 \pm 123.167$ & $0.13$          & \multirow{-3}{*}{3} \\ \hline \rowcolor{gray!15}
                                                             & high & $0.180 \pm 0.070$ & $1.163   \pm 0.517$   & $125.205 \pm 69.049$  & $2.838 \pm 0.064$ & $1204.315 \pm 2.236$   & $0.02$          &                     \\        \rowcolor{gray!15}
                                                             & mid  & $0.178 \pm 0.070$ & $1.385   \pm 0.370$   & $235.393 \pm 67.389$  & $2.551 \pm 0.055$ & $2328.240 \pm 45.730$  & $0.06$          &                     \\        \rowcolor{gray!15}
        \multirow{-3}{*}{\shortstack{\Sphincs+-192f\\(SHAKE)}} & low  & $0.209 \pm 0.115$ & $1.049   \pm 0.566$   & $230.640 \pm 71.643$  & $2.125 \pm 0.852$ & $9243.958 \pm 133.199$ & $0.04$          & \multirow{-3}{*}{3} \\ \hline
    \end{tabular}
\end{table*}

\section{Discussion}
\label{sec:discussion}

Based on the results presented in Section~\ref{sec:evaluation}, we can conclude that \ac{pqc} algorithms are feasible for implementation on lower-end and legacy embedded devices, including legacy \acp{ecu}. 
However, this comes with significant trade-offs regarding success rate, computational overhead, and security level.
For \ac{kem}, \textsc{Kyber}512 shows the lowest overhead (\SI{\approx 1.126}{\milli\second}) with a 59\% success rate in low-end settings, while higher-security alternatives like BIKE Level-5 and hqc-256 introduce prohibitive overheads (\qtyrange{\approx 1961}{4921}{\milli\second}).
Similarly, for \ac{dsa}, \textsc{Falcon}-512 achieves moderate overhead (\SI{\approx 167.53}{\milli\second}) with a 75\% success rate, whereas Dilithium Level 2 demonstrates lower overhead but limited success.
Thus, lightweight \ac{pqc} schemes may be viable with optimizations, but high-security implementations remain challenging.

\begin{formal}
\textbf{RQ1 Takeaway} -- Lower-end embedded devices can support \ac{pqc}, but only with low-overhead algorithms, requiring trade-offs in security or reliability and limiting their suitability for critical tasks.
\end{formal}

\ac{pqc} can also be implemented in time-sensitive applications, but their feasibility depends on the specific algorithm and security level.
For \ac{kem}, \textsc{Kyber}512 achieves a low overhead (\SI{\approx 1.1}{\milli\second}) with a high success rate (0.93), making it suitable for real-time systems when higher-end embedded devices are available.
In contrast, schemes like hqc-256 and BIKE Level-5 exhibit significantly higher overhead (\SI{< 250}{\milli\second} and \SI{> 1000}{\milli\second}, respectively), which may be impractical for strict timing constraints.
For \ac{dsa}, \textsc{Falcon}-512 maintains a low verification time (\SI{\approx 0.1}{\milli\second}) but suffers from high key generation latency, while Dilithium Level 2 offers a more balanced trade-off.

\begin{formal}
\textbf{RQ2 Takeaway} -- Low-latency \ac{pqc} algorithms like \textsc{Kyber}512 and Dilithium Level 2 can be deployed in time-sensitive applications when higher-end \acp{ecu} are available. Still, high-security schemes with excessive overhead may pose challenges.
\end{formal}
\ac{pqc} algorithms in embedded systems present a trade-off between computational overhead and security level.
Higher security levels, such as those in \textsc{Kyber}1024, BIKE Level-5, and hqc-256, require significantly more processing time for key generation, encapsulation, and decapsulation, leading to increased overhead.
Meanwhile, lower security levels, such as \textsc{Kyber}512 and \textsc{Falcon}-512, offer reduced computational costs but a lower success rate and resilience against quantum attacks.
Indeed, BIKE Level-5 exhibits high overhead (\SI{\approx 1961}{\milli\second}) but ensures high security, while \textsc{Kyber}512 maintains minimal overhead (\SI{\approx 1.1}{\milli\second}) at the cost of lower security.

\begin{formal}
\textbf{RQ3 Takeaway} -- Higher security in \acp{pqc} comes at the cost of significantly increased computational overhead in embedded devices, making it crucial to balance performance and security in embedded systems.
\end{formal}
\section{Conclusions}
\label{sec:conclusions}

This paper presented PQ-CAN, a novel framework for the simulation of \ac{pqc} algorithms in embedded systems and assessed their integration into automotive \acp{ecu} within simulated \ac{can} bus environments.
Results indicate that communication overhead is the primary performance bottleneck, with \textsc{Kyber} and \textsc{Falcon}-512 emerging as the most efficient key encapsulation and signature schemes, while Classic McEliece and \Sphincs+ proved impractical due to excessive execution times.
Future work should focus on real-world testing on physical \ac{can} networks, improving underperforming implementations, and exploring other \ac{pqc} schemes.
The framework could also be extended beyond automotive applications to other constrained environments, such as industrial control systems, avionics communication, or IoT networks relying on lightweight message protocols, broadening its applicability in post-quantum security.

\balance
\bibliographystyle{IEEEtran}
\bibliography{IEEEabrv,references}

\end{document}